\documentclass[12pt]{article}

\catcode`\@=11

\global\arraycolsep=2pt
\usepackage{amsbsy,amssymb,latexsym,amsfonts,amsmath}
\usepackage{graphicx,color}

\allowdisplaybreaks

\begin{document}
\begin{flushright}
\parbox{4.2cm}
{RUP-23-3}
\end{flushright}

\vspace*{0.7cm}

\begin{center}
{ \Large More on renormalizable exceptions to Nelson-Seiberg theorem}
\vspace*{1.5cm}\\
{Yu Nakayama and Takanobu Yoshida}
\end{center}
\vspace*{1.0cm}
\begin{center}

Department of Physics, Rikkyo University, Toshima, Tokyo 171-8501, Japan

\vspace{3.8cm}
\end{center}

\begin{abstract}
The Nelson-Seiberg theorem dictates conditions for the spontaneous breaking of the supersymmetry in Wess-Zumino models with generic, possibly non-renormalizable, superpotential; the existence of the R-symmetry is necessary while the spontaneous breaking of the R-symmetry is sufficient. If we restrict ourselves to generic but renormalizable theories, however, there exist Wess-Zumino models whose vacua break the R-symmetry spontaneously while preserving the supersymmetry. The classification and conditions of such renormalizable exceptions are under active study. We give some new examples of spontaneous breaking of the R-symmetry with preserved supersymmetry that are not covered in the literature. 
\end{abstract}

\thispagestyle{empty} 

\setcounter{page}{0}

\newpage

\section{Introduction}
Supersymmetry is a beautiful theoretical framework to unify matter and force, yet it has not been observed in particle physics experiments. If supersymmetry exists in fundamental physics, it must be spontaneously broken in the current universe. The spontaneous breaking of the supersymmetry in particle physics is attractive because it may explain a hierarchy problem as well as a dark matter problem in particle physics.

In many situations, even with the strong gauge dynamics involved, a study of the spontaneous breaking of supersymmetry can be reduced to a study of effective Wess-Zumino models. In a classic paper \cite{Nelson:1993nf}, Nelson and Seiberg gave the condition for the spontaneous breaking of the supersymmetry in Wess-Zumino models with generic, possibly non-renormalizable, superpotential; the existence of the R-symmetry is necessary while the spontaneous breaking of the R-symmetry is sufficient. 

Despite the beauty of the Nelson-Seiberg theorem, the assumption of the theorem may not be met in actual examples. If we restrict ourselves to generic but renormalizable theories, for instance, there exist ``generic" but renormalizable Wess-Zumino models whose vacua break the R-symmetry spontaneously while preserving the supersymmetry. This is because the genericity assumed in the Nelson-Seiberg theorem is stronger than the ``genericity" in {\it renormalizble} Wess-Zumino models. In addition, it would be preferable if we were able to know when the spontaneous breaking of the R-symmetry does or does not occur before solving vacuum equations or F-term equations. 

In a series of works \cite{Ray:2007wq}\cite{Komargodski:2009jf}\cite{Sun:2011fq}\cite{Kang:2012fn}\cite{Sun:2019bnd}\cite{Li:2020wdk}\cite{Amariti:2020lvx}\cite{Sun:2021svm}\cite{Li:2021ydn}\cite{Brister:2021xca}\cite{Brister:2022rrz}\cite{Brister:2022vsz} (see \cite{Sun:2022xdl} for the most recent review), they have studied the revised conditions for the spontaneous breaking of supersymmetry in ``generic" but renormalizable Wess-Zumino models. The central question is when the ``generic" superpotential in renormalizable Wess-Zumino models can be regarded as a generic superpotential in the sense of the Nelson-Seiberg theorem. In particular, they have given a sufficient condition for the existence of ``generic" but renormalizable Wess-Zumino models whose vacua break the R-symmetry spontaneously while preserving the supersymmetry. The condition turns out to be not necessary because they find another exceptional case in \cite{Brister:2022rrz} and we will show more such examples in this paper with systematic approaches to generate them.

The organization of the paper is as follows. In section 2, we review the Nelson-Seiberg theorem and its generalizations. In section 3, we give new examples of renormalizable exceptions  to the Nelson-Seiberg theorem. In section 4, we conclude the paper with some discussions.

\section{Spontaneous supersymmetry breaking in renormalizable Wess-Zumino models}
In this paper, we study  Wess-Zumino models with holomorphic superpotential $W(\Phi^i)$. We assume that the K\"ahler potential is everywhere regular, and the condition for the existence of the supersymmetric vacua is given by solving the F-term equations $\partial_{\Phi^i} W = 0$ for all chiral superfields $\Phi^i$. In this paper, we state that the supersymmetry is spontaneously broken when there is no solution to the F-term equations, and we will not commit ourselves to the study of the potential, which requires the knowledge of the effective K\"ahler potential.

Let us first review the Nelson-Seiberg theorem \cite{Nelson:1993nf}. It dictates conditions for the spontaneous breaking of the supersymmetry in Wess-Zumino models with generic, possibly non-renormalizable, superpotential; the existence of the R-symmetry is necessary while the spontaneous breaking of the R-symmetry is sufficient. The Nelson-Seiberg theorem itself, however, does not tell if the R-symmetry is spontaneously broken or not.

Let us now restrict ourselves to  renormalizable Wess-Zumino models, whose superpotential is up to cubic.  In \cite{Li:2021ydn} (called ``theorem 5" in \cite{Sun:2022xdl}), it was shown that in renormalizable Wess-Zumino models, R-symmetry is necessary for the spontaneous breaking of supersymmetry. Furthermore, with the existence of the R-symmetry, it was shown that it is necessary to have $N_2 > N_0$, where $N_2$ is a number of chiral superfields whose R-charge is two, and $N_0$ is a number of chiral superfields whose R-charge is zero. It means that it is sufficient to have $N_2 \le N_0$ to guarantee vacua with R-symmetry and supersymmetry both preserved.

The importance of the R-charge two chiral superfields comes from the fact that it can appear linearly in the superpotential. Without the linear term in the superpotential, one can easily convince ourselves that there is always a supersymmetric vacuum at the origin of the field space in renormalizable Wess-Zumino models.

The existence of the R-charge zero superfields, then effectively removes the linear term in the superpotential by acquiring the vacuum expectation values. This intuitively explains why the comparison between $N_2$ and $N_0$ is important.

These cases are consistent with the Nelson-Seiberg theorem.
Thus exceptions exist only when $N_2 > N_0$. We can argue that {\it if} the genericity assumption of the Nelson-Seiberg theorem had applied here, we would obtain vacua with R-symmetry and supersymmetry both broken \cite{Kang:2012fn} (called ``theorem 2" in \cite{Sun:2022xdl}), and in many examples, it is indeed the case. For example, the simplest O’Raifeartaigh model \cite{Shih:2007av}
\begin{align}
W = X(1+\tilde{P}\tilde{Q}) + \tilde{Q}A + \tilde{P}^2 \ . \label{Oraif}
\end{align}
with R-charge assignment $R(\{X,\tilde{P},\tilde{Q},A \}) = \{2,1,-1,3\}$
has $N_2 = 1$ and $N_0=0$, and it breaks both R-symmetry and supersymmetry spontaneously.

However, the genericity assumption in the sense of the Nelson-Seiberg theorem can be violated in ``generic" renormalizable Wess-Zumino models. At this point, we should emphasize again that the meaning of ``genericity" in the renormalizable Wess-Zumino model is that we include all the renormalizable terms (i.e. terms up to cubic) that are consistent with the R-symmetry, and this does not guarantee the genericity assumption made in the proof of the Nelson-Seiberg theorem.

In the proof of the Nelon-Seiberg theorem, we introduce the R-charge two (composite) chiral superfield $x$ and R-charge zero 
 (composite) chiral superfields $y_a$ to rewrite the superpotential as $W = x f(y_a)$. Under the assumption that $x$ has a vacuum expectation value, the F-term condition becomes $f=0$ and $\partial_{y_a} f = 0$, but we have one more constraint than variables, so we do not expect a solution for a generic holomorphic function $f$. In the renormalizable exceptions that we will discuss below, we effectively have no constant terms and no linear terms in $f$, which is regarded as a violation of the genericity assumption in the sense of the Nelson-Seiberg theorem and enables us to solve the F-term condition by setting $y_a=0$.

In \cite{Brister:2022vsz}, the brute force investigation has been performed with up to five chiral superfields. The result seems to suggest that order ten percent of ``generic" renormalizable Wess-Zumino models with $N_2 > N_0$
 are exceptions and possess vacua with the spontaneous breaking of R-symmetry while supersymmetry is preserved.\footnote{We should emphasize that we do not know how the ratio scales when we increase the number of superfields. Intuitively, the number of R-symmetric Wess-Zumino models and the supersymmetry breaking models show factorial growth (the latter having one smaller argument) and the ratio tends to zero.}  
One of the simplest examples is the Wess-Zumino model whose superpotential is 
\begin{align}
W = X(1 + PQ) + X^2 A + PA^2 \ .
\end{align} \label{exampleth4}
The R-charge assignment is 
$R(\{X,{P},{Q},A \}) = \{2,6,-6,-2\}$.
The moduli space of vacua is given by a complex hyperbola $PQ=-1$. The R-symmetry is spontaneously broken while the supersymmetry is preserved.

To explain such exceptions, they came up with a sufficient condition for vacua with R-symmetry and supersymmetry preserved in models with $N_2 > N_0$ \cite{Sun:2021svm}. In \cite{Sun:2022xdl} it is  called ``theorem 4".

Let $N_{PQ}$ be a number of independent products of chiral superfields $P$ and $Q$ that have opposite (non-zero) R-charges.
For the existence of R-symmetry breaking but supersymmetry preserving vacua, it is sufficient that $N_0 < N_2 \le N_0 + N_{PQ}$.
When we count $N_{PQ}$, $P$, and $Q$ should not have any mass terms and they should appear only linearly in cubic terms in the  superpotential.

In the above example \eqref{exampleth4}, we have $N_2=1, N_{PQ}=1, N_0=0$, so the theorem applies while in \eqref{Oraif}, we have $N_2=1$, $N_{PQ}=0$, $N_0=0$, and this theorem does not apply. Note that in the latter case, although we have superfields $\tilde{P}$ and $\tilde{Q}$ that have an opposite R-charge, they appear in the quadratic terms, so we do not count them in $N_{PQ}$.

Is this condition necessary? In \cite{Brister:2022rrz} they found one example with nine chiral superfields which does not satisfy the requirement but shows spontaneous R-symmetry breaking with the preserved supersymmetry. See example (3-0) in the next section.  In the rest of the paper, we are going to show more examples of this sort with fewer chiral superfields. Along the way, we will clarify the origin of such examples. 

The reason why we imposed the condition ``When we count $N_{PQ}$, $P$, and $Q$ should not have any mass terms and they should appear only linearly in cubic terms in the  superpotential." is that otherwise, typically, the moduli space of R-symmetry breaking vacua (such as $PQ=-1$ above) is lifted. Obviously, if the mass term demands $P=0$ or $Q=0$, it is not compatible with $PQ=-1$.
In the next section, we will critically address what we mean by ``typically" here.

\section{New examples}
A strategy to find exceptions to the Nelson-Seiberg theorem is to add terms in the superpotential  which violate the assumption, but at the same time, we combine the added terms with a new massless chiral superfield, which will be called $B$ in the following, and absorb the effect of the superpotential. In other words, we extend the moduli space of $PQ=-1$ in $\mathbb{C}^2$ to a complex curve in $\mathbb{C}^{2+N_B}$.

The following examples are listed in increasing order of $N_B$. The coefficients in the superpotential must be generic, but for the purpose of simplicity of presentation, we have set them to be unity. More details for finding these examples can be found in the Master's thesis of the second author \cite{Yoshida}.

(example 1-1)

The superpotential is given by
\begin{align}
W = X(1+ \tilde{P}\tilde{Q})+ A(\tilde{P}  + \tilde{Q} B) + X^2 A
\end{align}
with the R-charge assignment $R(\{X,\tilde{P},\tilde{Q},A,B\}) = \{2,4,-4,-2,8 \}$. The moduli space is a complex curve: $\tilde{P}\tilde{Q}=-1, B\tilde{Q}=-\tilde{P} $.  In this model, $N_2=1$, $N_0=0$, $N_{PQ}=0$ and it does {\it not} satisfy the condition $N_0 <N_2 \le N_0 + N_{PQ}$, but the R-symmetry is spontaneously broken with the supersymmetry preserved. We stress that $\tilde{P}$ and $\tilde{Q}$ have an opposite R-charge, but we do not count in $N_{PQ}$ because $\tilde{P}$ appears in the quadratic term. This is the reason why we put tilde here.

(example 1-2)

The superpotential is given by
\begin{align}
W = X(1 + \tilde{P}\tilde{Q}) + A( \tilde{P}  + \tilde{Q} B) + A^3
\end{align}
with the R-charge assignment $R(\{X,\tilde{P},\tilde{Q},A,B\}) = \{2,\frac{4}{3},-\frac{4}{3},\frac{2}{3},\frac{8}{3} \}$. The moduli space is a complex curve: $\tilde{P}\tilde{Q}=-1, B\tilde{Q}=-\tilde{P} $. The R-symmetry is spontaneously broken, but supersymmetry is preserved.

These two examples have only five chiral superfields so they must be on the list of \cite{Brister:2022vsz}. We believe there are no further exceptions in the list of \cite{Brister:2022vsz}, and the classification is now complete. 

(example 1-3)

The superpotential is given by
\begin{align}
W = X(1+  \tilde{P}\tilde{Q}) + A_1(\tilde{P}  + B^2) + X^2 A_2 + \tilde{P} A_2^2 
\end{align}
with the R-charge assignment $R(\{X,\tilde{P},\tilde{Q},A_1,A_2,B\}) = \{2,6,-6,-4,-2, 3 \}$. The moduli space is a complex curve: $\tilde{P}\tilde{Q}=-1, B^2=-\tilde{P} $.

(example 1-4)

The superpotential is given by
\begin{align}
W =  X(1+\tilde{P}\tilde{Q}) + A_1(\tilde{P} + \tilde{Q} B) + X^2 A_2 + \tilde{P} A_2 A_3 + A_3^2 + XA_1 A_3 + A_2^2 B
 \end{align}
with the R-charge assignment $ R(\{X,\tilde{P},\tilde{Q},A_1,A_2,A_3,B\}) = \{2,3,-3,-1,-2,1,6\}$. The moduli space is a complex curve: $\tilde{P}\tilde{Q}=-1, \tilde{Q}B=-\tilde{P} $. Note that in this model $A_1$ and $A_3$ have an opposite R-charge, so one may want to call them $\tilde{Q}'$ and $\tilde{P}'$. Since they have a quadratic superpotential, we still have $N_{PQ}=0$ anyway.

(example 1-5)

The superpotential is given by
\begin{align}
W = X(1 + \tilde{P}\tilde{Q}) + A_1 (\tilde{P}  + \tilde{Q} B ) + \tilde{P} A_2^2 + A_1^2 A_2 
\end{align}
with the R-charge assignment $R(\{X,\tilde{P},\tilde{Q},A_1,A_2,B\}) = \{2,\frac{6}{5},-\frac{6}{5},\frac{4}{5},\frac{2}{5}, \frac{12}{5} \}$. The moduli space is a complex curve: $\tilde{P}\tilde{Q}=-1, B \tilde{Q}=-\tilde{P} $.

More examples are available with $N_B >1$. The following examples are not exhaustive.

(example 2-1)

The superpotential is given by
\begin{align}
X(1+\tilde{P}\tilde{Q}) + A_1 (\tilde{P} + \tilde{Q}B_1) + A_2(\tilde{Q} + B_1 B_2) + X^2 A_1 + A_1^2 A_2 
\end{align}
with the R-charge assignment $R(\{X,\tilde
{P}
,\tilde{Q},A_1,A_2,B_1,B_2\}) = \{2,4,-4,-2,6,8,-12\}$. The moduli space is a complex curve: $\tilde{P}\tilde{Q}=-1,  \tilde{P} = -\tilde{Q} B_1, \tilde{Q}= -B_1 B_2 $.

(example 2-2)

The superpotential is given by
\begin{align}
W = X(1 + \tilde{P}\tilde{Q}) + A_1 (\tilde{P}  + \tilde{Q} B_1) + A_2(\tilde{Q}^2 + B_1 B_2) + X^2 A_1 
\end{align}
with the R-charge assignment $R(\{X,\tilde{P},\tilde{Q},A_1,A_2,B_1,B_2\}) = \{2, 4, -4,-2,10,8,-16 \}$. The moduli space is a complex curve: $\tilde{P}\tilde{Q} = -1, \tilde{P} = -\tilde{Q}B_1, \tilde{Q}^2 = B_1 B_2$.

(example 2-3)

The superpotential is given by
\begin{align}
W = X(1 + \tilde{P}\tilde{Q}) + A_1 (\tilde{P}  + \tilde{Q} B_1) + A_2(\tilde{Q}B_2 + B_1^2) + X^2 A_1 
\end{align}
with the R-charge assignment $R(\{X,\tilde{P},\tilde{Q},A_1,A_2,B_1,B_2\}) = \{2, 4, -4,-2,-14,8,20 \}$. The moduli space is a complex curve: $\tilde{P}\tilde{Q} = -1, \tilde{P} = -\tilde{Q}B_1, \tilde{Q} B_2 = - B_1^2$.

Let us consider $N_B=3$ cases, which include the original example (i.e. example 3-0) presented in \cite{Brister:2022rrz}.

(example 3-1)

The superpotential is given by
\begin{align}
W = X( 1+  \tilde{P}\tilde{Q}) + A_1 (\tilde{P} + \tilde{Q} B_1) + A_2(\tilde{Q} + B_1 B_2) + A_3 (B_2 + B_1 B_3) + X^2 A_1 + A_1^2 A_2
\end{align}
with the R-charge assignment $R(\{X,\tilde{P},\tilde{Q},A_1,A_2,A_3,B_1,B_2,B_3\}) = \{2,4,-4,-2,6,14,8,-12,-20 \}$. The moduli space is a complex curve: $\tilde{P}\tilde{Q} = -1,\tilde{P} = -\tilde{Q}B_1, -\tilde{Q}=B_1 B_2,B_3 B_1 = -B_2$.

(example 3-2)

The superpotential is given by
\begin{align}
W = X(1+\tilde{P}\tilde{Q}) + A_1 (\tilde{P} + \tilde{Q}B_1) + A_2 (\tilde{Q}^2 +B_1 B_2) + A_3(\tilde{Q} B_2 + B_1 B_3) + X^2 A_1 
\end{align}
with the R-charge assignment $R(\{X,\tilde{P},\tilde{Q},A_1,A_2,A_3,B_1,B_2,B_3\}) = \{2,4, -4,-2,10,22,8,-16,-28 \}$. The moduli space is a complex curve: $\tilde{P} \tilde{Q} = -1, \tilde{P} = -\tilde{Q}B_1, \tilde{Q}^2 = - B_1 B_2, \tilde{Q} B_2 = - B_1 B_3$.

(example 3-3)

The superpotential is given by
\begin{align}
W = X(1 +  \tilde{P}\tilde{Q}) + A_1(\tilde{P} + \tilde{Q} B_1) + A_2 (\tilde{Q}B_2 +  B_1^2)+ A_3 (B_2 + B_1 B_3) + X^2 A_1
\end{align}
with the R-charge assignment 
$R(\{X,\tilde{P},\tilde{Q},A_1,A_2,A_3,B_1,B_2,B_3\}) = \{2,4, -4,-2,-14,-18,8,20,12 \}$. The moduli space is a complex curve: $\tilde{P}\tilde{Q} = -1,\tilde{P} = -\tilde{Q}B_1, -\tilde{Q}B_2=B_1^2,B_1 B_3 = -B_2$.

(example 3-0)

This is the model found in \cite{Brister:2022rrz}; the superpotential is given by
\begin{align}
W = X(1+\tilde{P}\tilde{Q}) + A_1(B_1 + \tilde{P}^2) + A_2 (B_2 + B_3^2) + A_3(\tilde{P} + B_2^2 + B_1 \tilde{Q}) + A_1^2 B_3
\end{align}
with the R-charge assignment 
$R(\{X,\tilde{P},\tilde{Q},A_1,A_2,A_3,B_1,B_2,B_3\}) = \{2,\frac{8}{15}, -\frac{8}{15},\frac{14}{15},\frac{26}{15},\frac{22}{15},\frac{16}{15},\frac{4}{15},\frac{2}{15} \}$.

A comment about these new examples is in order. One may expect that adding non-renormalizable terms in the superpotential makes them compatible with the Nelson-Seiberg theorem. Interestingly, it is not always the case. Take example 1-1. One can see that adding {\it polynomial} non-renormalizable superpotential does not remove the R-symmetry breaking but the supersymmetry preserved vacuum that we obtained.

To see this structure, consider introducing the R-charge two (composite) chiral superfield $p = \tilde{P}^{\frac{1}{2}}$ and R-charge zero (composite) chiral superfields $\{x,q,a,b \} = \{ \frac{X}{\tilde{P}^\frac{1}{2}}, \tilde{Q}\tilde{P}, A\tilde{P}^{\frac{1}{2}}, \frac{B}{\tilde{P}^2}\}$. The superpotential becomes $W= p (x+xq + a + qba + a x^2)$, and it does not satisfy the genericity assumptions of the Nelson-Seiberg theorem. If we relax the renormalizability but still assume the polynomial superpotential in the original variable, the superpotential must be a polynomial $W = p f(x,q,a,b)$ but it is a special form: since $\tilde{P}$ must appear in a non-negative integer power, each term of $f$ must accompany at least one $x$ or $a$. This means that it is not a generic superpotential in the sense of the Nelson-Seiberg theorem because, without a constant term, $x=a=0$ can solve the F-term condition.


\section{Discussion}

In this paper, we have found new examples of Wess-Zumino models that can be regarded as exceptions to the Nelson-Seiberg theorem, which is not covered in the classification given in the literature. They first appear when the number of chiral superfields is  five. According to \cite{Brister:2022vsz}, with five chiral superfields, there exist 759 renormalizable Wess-Zumino models with the R-symmetry, among which 740 are compatible with the Nelson-Seiberg theorem. The remaining 19 are exceptions to the Nelson-Seiberg theorem and most of them are covered by the sufficient condition of \cite{Sun:2021svm}.\footnote{These numbers are updated in the newer version of \cite{Brister:2022vsz}, which takes account of our findings. We refer to the appendix of \cite{Brister:2022vsz} for the complete list.} We showed that there exist 2 
 renormalizable Wess-Zumino models that are not covered.

As we reviewed in section 2, when $N_2 \le N_0$, the ``generic" renormalizable Wess-Zumino models always have a vacuum with R-symmetry and supersymmetry preserved. By using a similar technique to the one used in section 3, however, we can obtain more subtle examples that may be regarded as an exception to the Nelson-Seiberg theorem. In these examples, in addition to an R-symmetry preserving vacuum, we also find vacua where R-symmetry is spontaneously broken while supersymmetry is preserved.\footnote{In the paragraph above, we have not included them as ``exceptions" because in \cite{Brister:2022vsz}, they are classified as theories with both R-symmetry and supersymmetry preserved (at a particular vacuum).} 

Let us present one such example. The superpotential is given by
\begin{align}
W = X(1+Y+Y^2 + \tilde{P}\tilde{Q}) + A (1+Y) \tilde{P} + X^2 A
\end{align}
with the R-charge assignment $R(\{X,Y,\tilde{P},\tilde{Q}, A \}) = \{2,0,4,-4,-2\}$. In addition to an R-symmetric and supersymmetry preserving vacuum $\tilde{P}=\tilde{Q}=0, Y = \frac{-1\pm \sqrt{3}i}{2}$, it has a moduli space of vacua, where R-symmetry is spontaneously broken but supersymmetry is preserved: $Y=-1,\tilde{P}\tilde{Q}= -1$, and $\tilde{Q}\neq 0, \tilde{P}=0, Y = \frac{-1+\sqrt{3}i}{2}$. The latter branch can be regarded as an exception to the Nelson-Seiberg theorem (i.e. the sufficiency condition of the spontaneous supersymmetry breaking) because the ``generic" renormalizable superpotential here is not generic in the sense of the Nelson-Seiberg theorem. More such examples can be found in the Master's thesis of the second author \cite{Yoshida}.

Including our examples, we have explained all the exceptions to the Nelson-Seiberg theorem in the renormalizable Wess-Zumino models up to five chiral superfields. Our technique to generate new examples is more or less systematic but still involves manual checks to see if the ``unwanted" superpotential terms, which will stabilize $B$, are compatible with the R-symmetry or not. We believe that our findings give an important clue to establish the ``necessary condition" for the R-symmetry broken but supersymmetry preserved vacua in 
 renormalizbale Wess-Zumino models. The establishment of the necessary condition means that we have the sought-after criterion to find spontaneous breaking of the supersymmetry without solving the F-term equations.

\section*{Acknowledgements}
We would like to thank Z.~Sun for the correspondence.
This work is in part supported by JSPS KAKENHI Grant Number 21K03581.

\end{document}